# The Globalization-Inequality Nexus: A Comparative Study of Developed and Developing Countries

Md Shah Naoaj
[1](Graduate Student, Economics Department, New York University, USA)

***Abstract:*** *This study examines the relationship between globalization and income inequality, utilizing panel data spanning from 1992 to 2020.Globalization is measured by the World Bank's global-link indicators such as FDI, Remittance, Trade Openness, and Migration while income inequality is measured by Gini Coefficientand the median income of 50% of the population. The fixed effect panel data analysis provides empirical evidence indicating that globalization tends to reduce income inequality, though its impact varies between developed and developing countries. The analysis reveals a strong negative correlation between net foreign direct investment (FDI) inflows and inequality in developing countries, while no such relationship was found for developed countries.The relationship holds even if we consider an alternative measure of inequality. However, when dividing countries by developed and developing groups, no statistically significant relationship was observed.Policymakers can use these findings to support efforts to increase FDI, trade, tourism, and migration to promote growth and reduce income inequality.*
***Key Word****: Globalization, Income Inequality, FDI, GiniCoefficient, Developing Countries, Developed Countries*

## I. Introduction

Globalization is quantified by the "Global Connectedness Index" through its examination of trade, capital, information, and people flows(Altman & Bastian, 2021). Due to globalization, global interconnectedness has been increasing in the last two decades in the form of more financial flows, foreign direct investments, trade in goods and services, and movements of people among the countries. The global links enable the economies of individual countries to grow and expand. Moreover, as national economies develop, their links expand and grow more complex. But does this global link influences income inequality? If it does, then does it influence developed and developing countries differently? This paper aims to examine the relationship between World Bank's global-link indicators, including Foreign Direct Investment, Remittances, Trade Openness, and Migration, and the income inequality indicator represented by the Gini Coefficient.

## II. Literature Review

The relationship between globalization and economic growth and other macroeconomic indicators has received much attention since the world saw a major trade liberalization process beginning in 1990. There is substantial literature that shows that there is an impact of foreign direct investment and economic growth (Borensztein, De Gregorio and Lee, 1998; Herzer, Klasen and Nowak-Lehmann, 2008; De Vita and Kyaw, 2009). But globalization doesn't constitute only FDI, many other economic factors are also contributing to the change of the shape of world poverty,development and inequality. Specifically, the causal relationship between the global connectedness indicators and income inequality was not received much attention until 2000 due to lack of data and sound economic theory.

Two channels that impact inequality due to FDI was suggested by Jensen and Rosas (2007). They argued that FDI brings capital into a country and reduces capital return but increases labor return as FDI competes with local capital to get the local workers, raising wages and reducing firms' profitability. It means that FDI reduces income inequality by reducing the wage gap. On the contrary, the authors argued that the FDI ideally does not attract all labor force but skilled workforces. Consequently, inequality increases due to the wage gap between skilled and unskilled workers. According to Velde (2003), there are three possible channels through which wage inequality in developing countries has been affected by FDI. First, a "composition effect" results from the fact that foreign firms tend to set up more intensive sectors of skilled labor, thus improving the position of these workers relative to the unskilled (Feenstra and Hanson, 1997). Second, FDI can affect the supply of skilled workers via training and specific contributions to general education (knowledge transfer).

Lastly, as advanced by Berman, Bound and Machin (1998), FDI can probably induce faster labor productivity growth both in foreign firms (technology transfer) and in local ones (secondary effects), and if productivity growth is skewed towards skilled sectors, then the gap between the sectors will grow. According to the World Economic Outlook (IMF) Chapter 4, 2007, the impact of globalization on inequality was limited, as trade globalization was found to decrease inequality, but financial globalization and foreign direct investment specifically were found to increase inequality. Technological progress was found to have a greater impact on inequality within countries.

Given the relevance of this topic, many empirical studies have been conducted on the FDI and remittance inflow on income inequality. However, no such studies have been done on based on the World Bank's global-link indicators[1]. This paper aims to this causal relationship between globalization and inequality by analyzing global link indicators as a proxy of globalization. The motivation for using global link indicators as a proxy for globalization stems from the innovative DHL Global Connectedness Index published jointly by DHL and New York University. The DHL Global Connectedness Index (GCI) is a research study that measures the extent of cross-border flows of trade, capital, information, and people. The study provides an analysis of the interconnectedness of countries around the world and highlights the most connected countries and regions. The GCI is widely cited in academic and policy circles as a valuable source of information on globalization and its impact on the global economy.

### III. Data And Methods

The purpose of this study is to investigate the correlation between globalization and income inequality by analyzing several key variables, including the Gini coefficient (gini), net foreign direct investment inflow (l_fdi_net_in), net remittance inflow (inward_remittance), trade openness (trade_gdp), which is represented as the total trade as a percentage of Gross Domestic Product, and tourism receipts (tourist_receipt). To control for any unobserved heterogeneity, additional relevant variables have been included. The data used in this study was sourced from the World Development Indicators, World Bank, covering the period from 1992 to 2020. Due to many missing variables, the data set is unbalanced. The summary statistics are given in Table 1.

**Table 1: Summary Statistics**

| Variable | Obs | Mean | Std. Dev. | Min | Max |
|---|---|---|---|---|---|
| gini | 1643 | 38.442 | 9.069 | 20.7 | 65.8 |
| pop_below_50 | 1598 | 14.087 | 5.944 | .9 | 34.6 |
| l_fdi_net_in | 4676 | 19.948 | 2.815 | 2.303 | 27.322 |
| tourist_receipt | 3494 | 5.237e+09 | 1.603e+10 | 100000 | 2.420e+11 |
| trade_gdp | 4617 | 80.505 | 44.579 | .021 | 380.104 |
| inward_remittance | 4573 | 4.249 | 8.221 | 0 | 167.432 |

To understand the effect of globalization on income inequality, panel data regression is used for both developed and developing countries with the following specifications. In specification one, Gini coefficient is used as the dependent variable and the other global-link indicators as independent variables.

$$gini_{it} = \alpha + \beta 1 l\_fdi\_net\_in_{it} + \beta 2 toursim\_receipt_{it} + \beta 1 trade\_gdp_{it} + \beta 1 inward\_remittance_{it} + Ui + e_{it}$$

In the second specification, a proxy variable for income inequality is used. The variable represents the percentage of the population living below 50% of the median income or consumption per capita.

$$pop\_bleow\_50\_percent_{it} = \alpha + \beta 1 l\_fdi\_net\_in_{it} + \beta 2 toursim\_receipt_{it} + \beta 1 trade\_gdp_{it} + \beta 1 inward\_remittance_{it} + Ui + e_{it}$$

The term Ui represents fixed effects by country in the equation, and $e_{it}$ is the error term. The definitions of each of the variable are given in annex 1.

To understand the impact ofglobalization on income inequality, at the first step, countries are divided into developed and developing/emerging countries according to the definition of IMF. Thepanel regression analysis is conducted using the Gini coefficient as a dependent variable and global link indicators as independent variables. In step 2, the impact is analyzed for developed and developing countries using percentage of population living below the median 50 percent of income of each country. In step 3, the results betweenthe developed and developing effectare compared.

---

[1] According to World Bank, the Global Links indicators give an overview of the economic growth and expansion enabled by the flows and connections between the world's economy and the economies of individual countries. These indicators assess the magnitude and direction of these flows, and record government actions such as tariffs, trade facilitation, and aid.

Fixed-effect panel regression is used to analyze the causality as panel data analysis eliminates the effects of time-invariant unobservable heterogeneity and provides better causal inference, improved efficiency, and reduced omitted variable bias. The choice between fixed-effect and random-effect models depends on the nature of the sample. A random-effect model is suitable when the entities in the sample are randomly selected from the population, while a fixed-effect model is appropriate when the sample entities effectively make up the entire population. The Hausman test was conducted to determine the preferred model, and the results rejected the null hypothesis, indicating that the fixed-effect model was preferred over the random-effect model. As a result, the fixed-effect model was used in this analysis.

## IV. Result

Table 2 shows the regression output of the causal relationship between globalization on income inequality. This table presents the results of a regression analysis of the relationship between income inequality, as measured by the Gini coefficient, and various indicators of globalization and control variables. The Gini coefficient measures the distribution of income within a country, with higher values indicating greater income inequality.The first two columns of the table show the results of regression models that include all 146 countries in the sample. The next two columns show the results of regression models that include only developing countries, while the last two columns show the results of regression models that include only developed countries.

The independent variables in columns 1, 3, and 5 are globalization indicators, namely net foreign direct investment (FDI), tourist receipts, trade-to-GDP ratio, inward remittances, and the population growth rate. Meanwhile, columns 2, 4, and 6 also include control variables such as the primary education rate and GDP growth rate for robustness checks.

**Table 2: Effect of globalization on income inequality (developing vs developed)**

| VARIABLES | (1) gini (all countries) | (2) gini (all countries) | (3) gini (developing countries) | (4) gini (developing counties) | (5) gini (developed countries) | (6) gini (developed countries) |
|---|---|---|---|---|---|---|
| l_FDI_net_in | -0.616*** | -0.462*** | -0.672*** | -0.587*** | -0.097 | -0.069 |
|  | (0.170) | (0.171) | (0.202) | (0.202) | (0.158) | (0.215) |
| tourist_receipt | -0.000 | -0.000 | -0.000*** | -0.000*** | 0.000** | 0.000 |
|  | (0.000) | (0.000) | (0.000) | (0.000) | (0.000) | (0.000) |
| trade_gdp | 0.017 | 0.014 | 0.019 | 0.014 | 0.002 | -0.003 |
|  | (0.014) | (0.012) | (0.019) | (0.016) | (0.010) | (0.012) |
| inward_remittance | -0.120*** | -0.070 | -0.114*** | -0.059 | 0.212 | 0.167 |
|  | (0.040) | (0.045) | (0.040) | (0.048) | (0.323) | (0.283) |
| gdp_growth |  | 0.010 |  | 0.029 |  | -0.052 |
|  |  | (0.032) |  | (0.036) |  | (0.052) |
| toatal_pop |  | -0.000 |  | 0.000 |  | -0.000 |
|  |  | (0.000) |  | (0.000) |  | (0.000) |
| primary_edu_rate |  | -0.049* |  | -0.047* |  | 0.018 |
|  |  | (0.025) |  | (0.027) |  | (0.064) |
| Constant | 51.161*** | 53.033*** | 55.324*** | 57.127*** | 33.300*** | 32.829*** |
|  | (3.413) | (5.040) | (4.040) | (5.663) | (3.854) | (7.917) |
|  |  |  |  |  |  |  |
| Observations | 1,209 | 863 | 870 | 646 | 339 | 217 |
| R-squared | 0.074 | 0.094 | 0.126 | 0.125 | 0.014 | 0.028 |
| No of country | 146 | 129 | 114 | 103 | 32 | 26 |

Standard errors in parentheses (*** p<0.01, ** p<0.05, * p<0.1)

For all countries, net foreign direct investment has a statistically significant negative relationship with income inequality, with a coefficient of -0.616. This suggests that an increase in net foreign direct investment is associated with a decrease in income inequality. The relationship holds when we consider related control variables. Meanwhile, tourist receipts and trade to GDP ratio have no statistically significant relationship with income inequality. Inward remittances have a statistically significant negative relationship with income inequality, with a coefficient of -0.120. GDP growth rate is also positively related to income inequality, with a coefficient of 0.010. The primary education rate has a statistically significant negative relationship with income inequality, with a coefficient of -0.049.

When looking at the results for developing countries, we find similar results. Net foreign direct investment has a statistically significant negative relationship with income inequality, with a coefficient of -0.672. Tourist receipts have a statistically significant negative relationship with income inequality. Inward remittances have a statistically significant negative relationship with income inequality, with a coefficient of -0.114. GDP growth rate is negatively related to income inequality, with a coefficient of -0.052. The primary education rate has a statistically significant positive relationship with income inequality, with a coefficient of 0.018.

For developed countries, the results are different. Net foreign direct investment has a positive but not statistically significant relationship with income inequality, with a coefficient of -0.097. Tourist receipts have a positive but not statistically significant relationship with income inequality, with a coefficient of 0.000. Inward remittances have a statistically significant positive relationship with income inequality, with a coefficient of 0.212. GDP growth rate has no statistically significant relationship with income inequality. The primary education rate has a positive but not statistically significant relationship with income inequality, with a coefficient of -0.047.

In general, the results suggest that globalization, as measured by net foreign direct investment and inward remittances, is associated with lower levels of income inequality in both all countries and developing countries. However, trade as a percentage of GDP has a positive effect on income inequality in developing countries, but this effect is not statistically significant in developed countries. In developed countries, the relationship between globalization and income inequality is less clear. The primary education rate appears to be negatively related to income inequality in all countries, suggesting that an increase in education may help reduce income inequality. These findings offer insights into the relationship between globalization and income inequality and provide a foundation for further research on the topic.

The number of observations in the models ranges from 646 in the developing countries model to 1,209 in all countries model. The R-squared values indicate that the models explain a low to moderate amount of variation in the dependent variable. The number of countries in the sample ranges from 103 in the developing countries model to 146 in the all-countries model.The results of this analysis should be interpreted with caution, as they are based on cross-country comparisons and do not account for country-specific factors that may influence income inequality.

## V. Alternative measures of Inequality

Table3 shows an alternative measure of inequality which is the percentage of the population living below 50% of median income. The results from the regression analysis show that there is a negative relationship between the population below the poverty line of 50 and the amount of foreign direct investment (FDI) received (l_fdi_net_in), with the coefficient being statistically significant at the 1% level (p-value < 0.01). This result holds for all countries and developing countries.

The trade openness measured by trade to GDP ratio (trade_gdp) has a positive relationship with poverty, with a statistically significant coefficient of 0.0195 (p-value < 0.05) for all countries. However, this relationship is not significant for either developed or developing countries separately.Inward remittances (inward_remittance1) have a negative relationship with poverty, but the coefficient is only statistically significant at the 10% level (p-value < 0.1) for developing countries (-0.055). This relationship is not significant for developed countries.

Primary education rate (primary_edu_rate) has a negative relationship with poverty, with the coefficient being statistically significant for both developing and developed countries. The coefficient is larger for developing countries (-0.0233) compared to developed countries (-0.0208).The R-squared value, a measure of goodness of fit, is higher for developing countries (0.129) compared to developed countries (0.027). This suggests that the model fits better for developing countries than for developed countries.

The results suggest that FDI has a negative relationship with poverty in both developing and developed countries, with the relationship being stronger in developing countries. Other variables such as trade openness, inward remittances, and primary education rate also have a negative relationship with poverty, but their coefficients are not statistically significant for all countries or for either developing or developed countries. The results of this analysis can be used to inform policies aimed at reducing poverty in both developing and developed countries.

**Table 3: Alternative measure of inequality**

| VARIABLES | (1) pop_below_50 All countries | (2) pop_below_50 All countries | (3) pop_below_50 developing countries | (4) pop_below_50 developing countries | (5) pop_below_50 developed countries | (6) pop_below_50 developed countries |
|---|---|---|---|---|---|---|
| l_fdi_net_in | -0.433*** | -0.316*** | -0.465*** | -0.374*** | -0.124 | -0.113 |
|  | (0.0932) | (0.0941) | (0.109) | (0.109) | (0.122) | (0.125) |
| tourist_receipt | -0 | -0 | -1.17e-10** | -1.00e-10* | 0 | 0** |
|  | (0) | (0) | (0) | (5.47e-11) | (0) | (0) |
| trade_gdp | 0.0195* | 0.0108 | 0.0205 | 0.0102 | 0.0118 | -0.00177 |
|  | (0.0101) | (0.00953) | (0.0138) | (0.0125) | (0.00759) | (0.0106) |
| inward_remittance1 | -0.0601* | -0.0260 | -0.0550 | -0.0207 | -0.0314 | -0.117 |
|  | (0.0354) | (0.0470) | (0.0359) | (0.0481) | (0.311) | (0.303) |
| gdp_growth1 |  | -0.0145 |  | 0.00103 |  | -0.0677** |
|  |  | (0.0305) |  | (0.0372) |  | (0.0296) |
| toatal_pop1 |  | -8.93e-08* |  | -6.85e-08 |  | 4.77e-08 |
|  |  | (4.66e-08) |  | (4.26e-08) |  | (1.71e-07) |
| primary_edu_rate1 |  | -0.0233 |  | -0.0201 |  | -0.0208 |
|  |  | (0.0174) |  | (0.0180) |  | (0.0420) |
| Constant | 22.19*** | 24.57*** | 24.48*** | 26.57*** | 12.14*** | 13.37** |
|  | (1.930) | (2.699) | (2.224) | (2.940) | (3.102) | (6.100) |
| Observations | 1,188 | 849 | 849 | 632 | 339 | 217 |
| R-squared | 0.077 | 0.102 | 0.129 | 0.131 | 0.027 | 0.096 |
| Number of country | 143 | 128 | 111 | 102 | 32 | 26 |

Robust standard errors in parentheses
*** p<0.01, ** p<0.05, * p<0.1

## VI. Conclusion

The study concludes that globalization has a positive effect on reducing income inequality, but its impact varies between developed and developing countries. The results show that net foreign direct investment inflow has a statistically significant negative impact on inequality for all countries and specifically for developing countries. However, no such relationship was found for developed countries even when analyzing the data over a shorter time frame. These findings suggest that globalization can not only enhance economic growth but also reduce income inequality in developing countries. The alternative measure of inequality supports these conclusions, but without a statistically significant difference between developed and developing countries. Given these outcomes, policymakers should be proactive in promoting globalization through the promotion of foreign direct investment, tourism, trade, and mobility to drive economic growth and decrease inequality.The findings offer insights into the relationship between globalization and income inequality and poverty, but it is important to keep in mind that the results should be interpreted with caution as they are based on cross-country comparisons and do not account for country-specific factors that may influence these outcomes.